\documentclass{jfm}
\usepackage{natbib}
\usepackage{amsbsy}
\usepackage{psfrag}
\usepackage{amsmath}
\usepackage{soul}
\usepackage{lineno}
\usepackage[dvipsnames]{xcolor}
\usepackage{mathrsfs} 
\usepackage{amssymb}
\usepackage{mathtools}
\usepackage{graphicx}
\usepackage{epstopdf,epsfig}
\usepackage{comment}
\usepackage{multirow}

\usepackage[dvipsnames]{xcolor} % Added for colored comments used in drafting the paper.
\usepackage[hidelinks]{hyperref} % Added for easy navigation of pdf.

\newcommand{\blue}{\color{black}}
\newcommand{\bluenew}{\color{black}}

\newcommand{\black}{\color{black}}

\shorttitle{POD-mode-augmented wall model for LES}
\shortauthor{C. Hansen, X. Yang and M. Abkar}

\title{\Large{POD-mode-augmented wall model and its applications to flows at non-equilibrium conditions}}

\author{\normalsize{Christoffer Hansen\aff{1},
Xiang I. A. Yang\aff{2}
\corresp{\email{xzy48@psu.edu}}
 \and Mahdi Abkar\aff{1} \corresp{\email{abkar@mpe.au.dk}}}}

\affiliation{
\aff{1}Department of Mechanical and Production Engineering, Aarhus University, 8200 Aarhus N, Denmark
\aff{2}Department of Mechanical Engineering, Pennsylvania State University, State College, PA, 16802, USA
}
\begin{document}
\maketitle
%%
%%
%%\linenumbers
\begin{abstract}
Insights gained from modal analysis are invoked for predictive large-eddy simulation (LES) wall modeling.
Specifically, we augment the law of the wall (LoW) by an additional mode based on a one-dimensional proper orthogonal decomposition (POD) applied to a 2D turbulent channel.
The constructed wall model contains two modes, i.e., the LoW mode and the POD-based mode, and the model matches with the LES at two, instead of one, off-wall locations.
To show that the proposed model captures non-equilibrium effects, we perform {\it a-priori} and {\it a-posteriori} tests in the context of both equilibrium and non-equilibrium flows.
The {\it a-priori} tests show that the proposed wall model captures extreme wall-shear stress events better than the equilibrium wall model.
The model also captures non-equilibrium effects due to adverse pressure gradients.
The {\it a-posteriori} tests show that the wall model captures the rapid decrease and the initial decrease of the streamwise wall-shear stress in channels subjected to suddenly imposed adverse and transverse pressure gradients, respectively, both of which are missed by currently available wall models.
These results show promise in applying modal analysis for turbulence wall modeling.
In particular, the results show that employing multiple modes helps in the modeling of non-equilibrium flows.
\end{abstract}
\begin{keywords}
...
\end{keywords}
\section{Introduction}
\begin{figure}
\centering
\includegraphics[width=0.9\textwidth]{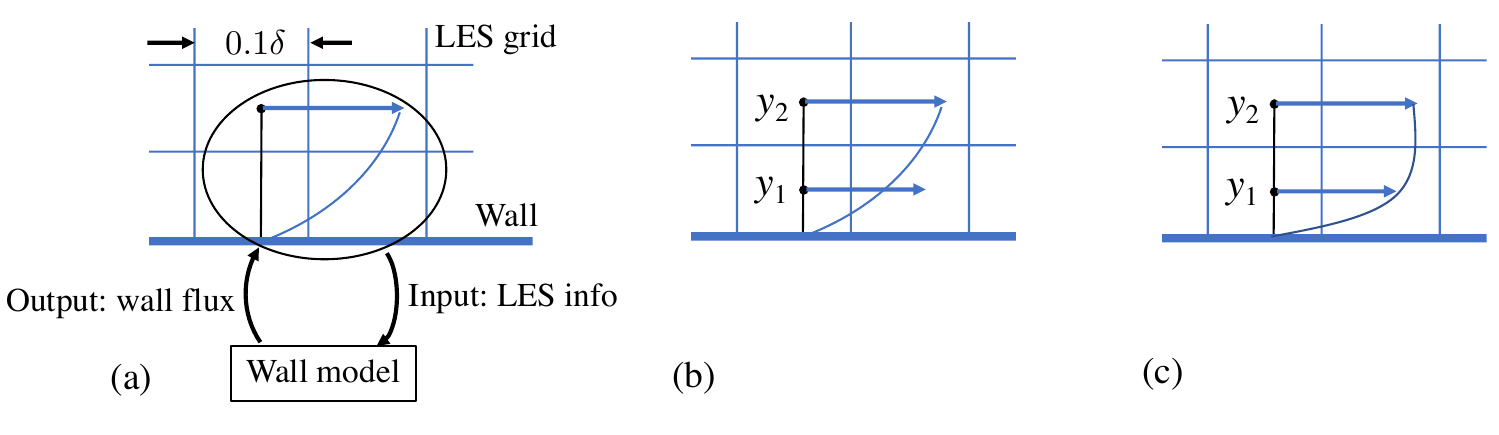}
\caption{\label{fig:sketch-WM} (a) Schematic of WMLES. 
(b) Schematic of the equilibrium wall model (EWM). 
Here, the wall model (WM) and the LES match at the second off-wall grid point. 
(c) Schematic of the POD-augmented WM. 
The WM and the LES match at two off-wall grid points.} 
\end{figure}

The strict near-wall grid resolution requirements for large-eddy simulations ({\blue LES}) make wall modeling a necessity at high Reynolds numbers \citep{choi2012grid,yang2021grid}.
The basic concept of wall-modeled LES (WMLES) is shown in schematic form in figure \ref{fig:sketch-WM}(a). 
As the LES grid in WMLES is coarse and scales with the boundary-layer thickness, the wall-shear stress and wall-heat flux cannot be computed per the discretization scheme.
Instead, a wall model is employed.
It takes LES information in the wall-adjacent cell(s) and computes the wall-shear stress and wall-heat flux.
These wall fluxes are then used as Neumann boundary conditions at the wall in the LES.

The most extensively used type of wall model is the equilibrium wall models (EWMs) {\blue\citep{schumann1975subgrid,bou2005scale,kawai2012wall, de2021unified}}.
The algebraic variant of EWMs computes the wall fluxes according to some mean flow scaling in the wall-adjacent computational cell(s) which is usually taken as the law of the wall (LoW).
The mean flow scaling is matched with the LES velocity at an off-wall location locally and instantaneously, resulting in algebraic equations for the friction velocity and the friction temperature.
The wall fluxes are then computed based on these friction quantities.
Another popular variant of EWMs is based on solving the thin boundary-layer equations (TBLEs).
In these models, only the Reynolds stress term and the viscous stress term are retained in the TBLEs, and the Reynolds stress term is closed using a zero-equation eddy viscosity model \citep{kawai2012wall,yang2018semi,chen2022wall}.
The resulting ordinary differential equations (ODEs) are then solved on a 1D fine near-wall grid with the off-wall boundary condition provided by matching with the LES in the wall-adjacent cell(s).
The wall fluxes can then be calculated from the ODE solutions on the wall model grid. 

Although the EWMs discussed above have much in their favor regarding simplicity, model stability, and a low computational cost, it has long been known that EWMs struggle in non-equilibrium flows, and thus, improvements in wall modeling beyond the EWMs are needed \citep{piomelli2002wall,larsson2015large,bose2018wall}. 
In the following, we review a few recent approaches.
One approach is based on solving the full TBLEs with all of the non-equilibrium terms included \citep{park2014improved}.
This method, however, comes at the cost of solving a full set of partial differential equations (PDEs), which often adds a 100\% overhead to the LES as in \cite{park2016numerical}.
Approaches that account for non-equilibrium effects at a lower computational cost are the integral wall model (IWM) by \cite{yang2015integral} and the Lagrangian relaxation towards equilibrium wall model (LaRTE) by \cite{fowler2022lagrangian}.
Both models perform wall-normal integration of the full TBLEs with an assumed analytical form for the LES-grid filtered velocity.
This results in a wall model that only requires the solution of algebraic equations.
Another interesting wall modeling approach is the dynamic slip wall model \citep{bose2014dynamic,bae2019dynamic}.
In this approach, a slip boundary condition is derived from the filtered Navier-Stokes equations, which is then used instead of the traditional Neumann boundary condition at the wall. 

\begin{figure}
\centering
\includegraphics[width=0.8\textwidth]{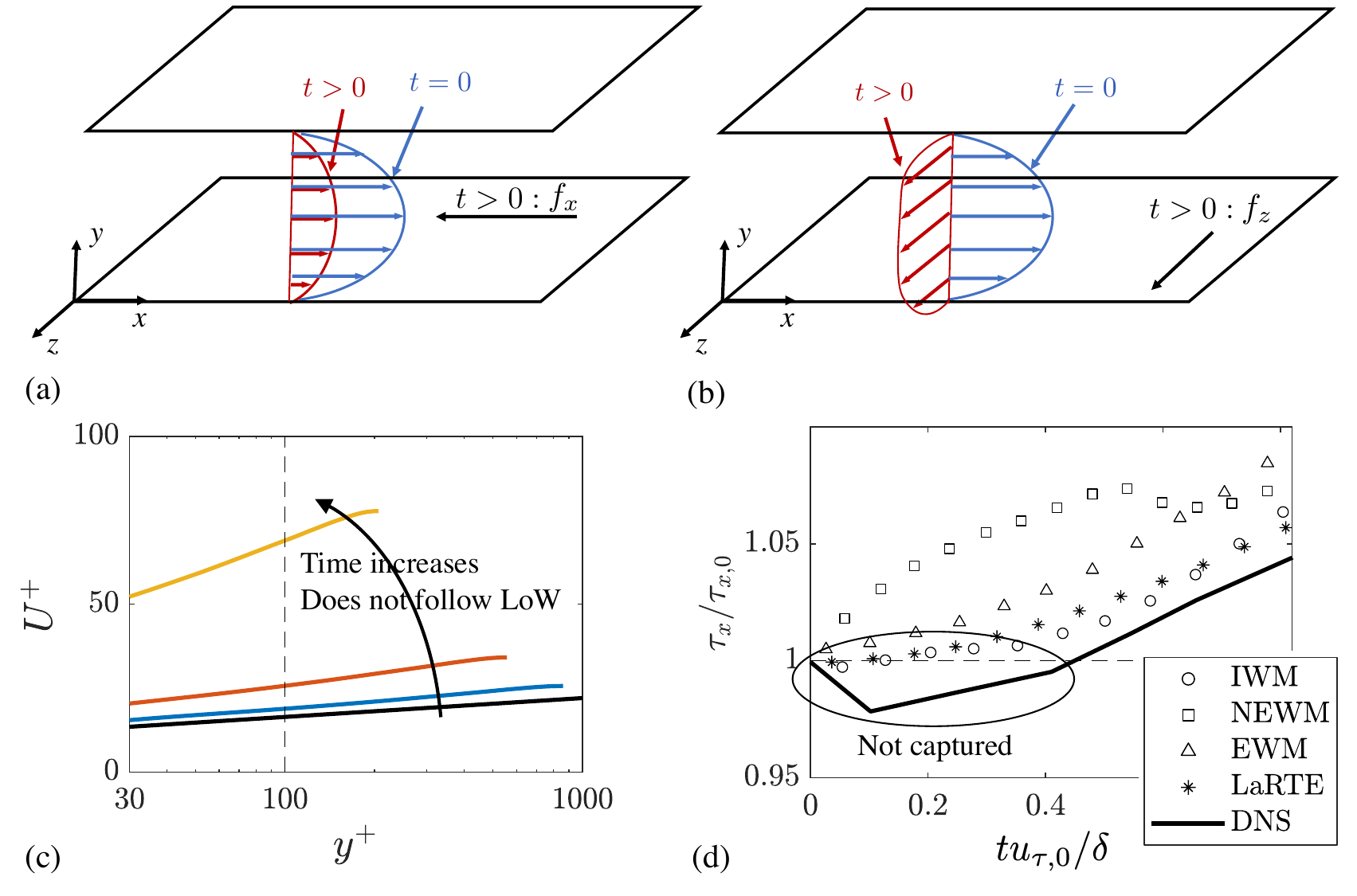}
\caption{\label{fig:sketch-chan} Schematics of a 2D channel subjected to (a) a suddenly imposed APG and (b) a suddenly imposed TPG at $t=0$.
Here, $f_x$ and $f_z$ are the imposed pressure gradients in the $x$ and $z$ directions, $t$ is the time, $x$, $y$, and $z$ are the three Cartesian coordinates.
(c) Evolution of the velocity profiles in time after an APG $f_x=100 f_{x,0}$ is suddenly applied to a $Re_\tau=1000$ channel.
The solid black line corresponds to the log law $U^+=\log(y^+)/\kappa+B$, where $\kappa=0.4$ and $B=5$.
Here, $u_\tau$ is the friction velocity, $f_{x,0}$ is the driving force of the 2D channel, $U$ is the streamwise velocity, and the superscript $+$ denotes normalization by wall units.
(d) Evolution of the $x$-direction wall-shear stress after a TPG $f_z=10f_{x,0}$ is applied to a $Re_\tau=1000$ channel.
DNS: direct numerical simulation \citep{lozano2020non}, IWM: integral wall model \citep{yang2015integral}, NEWM: non-equilibrium wall model \citep{park2014improved}, EWM: equilibrium wall model \citep{yang2017log}, and LaRTE: Lagrangian relaxation towards equilibrium wall model \citep{fowler2022lagrangian}. 
}
\end{figure}

We consider boundary layers subjected to adverse and transverse pressure gradients, which remain challenges for wall modeling \citep{bose2018wall}. 
Adverse pressure gradients (APGs) and transverse pressure gradients (TPGs) arise in many flows \citep{monty2011parametric,volino2020non,goc2021large}.
For the present work, we focus on the model problem shown in figure \ref{fig:sketch-chan}(a, b) where a 2D channel is subjected to a suddenly imposed pressure gradient \citep{na1998direct,he2015transition,lozano2020non}, here, a suddenly imposed APG or TPG.
The flow decelerates in (a) and changes direction in (b).
Relevant to near-wall turbulence modeling efforts is the behavior of the mean flow, which is the input to a wall model, and the wall-shear stress, which is the output of a wall model.
Figure \ref{fig:sketch-chan}(c) shows the evolution of the mean flow in (a) after an APG $f_x=100f_{x,0}$ is suddenly imposed to a $Re_\tau=1000$ channel.
{\blue Here $Re_\tau = u_\tau \delta / \nu$ with $u_\tau$ the friction velocity, $\delta$ is the half-channel width, and $\nu$ is the kinematic viscosity.}
The mean velocity profile in inner units is above the LoW, and if one were to apply the LoW to predict the wall-shear stress, the wall-shear stress would be grossly over-predicted.
Figure \ref{fig:sketch-chan}(d) shows the evolution of the $x$-direction wall-shear stress, $\tau_x$, after a TPG $f_z=10f_{x,0}$ is applied to a $Re_\tau=1000$ channel.  
The wall-shear stress decreases initially and then increases.
This behavior, however, is not captured by the available wall models.

This work aims to advance WMLES by accounting for the aforementioned non-equilibrium effects.
The idea is to augment the LoW such that the ansatz used for velocity reconstruction in the wall-adjacent cell(s) provides a more realistic description of the LES-grid-filtered flow than the LoW.
{\blue First, however, we briefly review algebraic EWMs based on the LoW to establish some context.
In these models, the following ansatz is used for the near-wall flow
\begin{equation}
    {\bf U} = {\bf c}{\rm LoW}(y^+) \, , 
    \label{eq:EWM}
\end{equation}
where ${\bf U} = (U,W)$ are the wall-parallel velocities at a distance $y$ from the wall, ${\rm LoW}(y^+)$ is usually taken to be the logarithmic law of the wall, and $y^+ = (u_\tau/\nu) y$ is the wall-unit-scaled distance from the wall.
We will take ${\rm LoW}(y^+)$ to include the viscous sublayer, the buffer layer, and the logarithmic layer, but not the wake layer.
Further, ${\rm LoW}(y^+)$ is taken to be nondimensional such that the friction velocity $u_\tau$ has been absorbed into the coefficients.
We propose to generalize these models by employing the augmented ansatz
\begin{equation}
    {\bf U}={\bf c}_{1}{\rm LoW}(y^+)+ {\bf c}_{2}g(y^+) \, ,
    \label{eq:WM}
\end{equation}
where $g(y^+)$ is, at this point, a generic nondimensional function.}
Figure \ref{fig:sketch-WM}(b, c) is a schematic of the velocity reconstruction according to \eqref{eq:EWM} and \eqref{eq:WM}, respectively.
Inclusion of the function $g$ gives the wall model in figure \ref{fig:sketch-WM}(c) the ability to account for deviations from the LoW scaling, and thus, {\blue to respond to both instantaneous fluctuations and non-equilibrium effects.
For convenience, we will refer to both of these cases as capturing non-equilibrium effects}.
\cite{yang2015integral} and \cite{lv2021wall} argued that $g\sim y$ to leading order, but other choices that more closely mimic the physics of {\blue wall-bounden turbulent flows} could yield better results.
To that end, we propose to obtain $g$ from modal analysis of {\blue the near-wall region of a 2D turbulent channel as a starting point for such investigations}.
While many possible choices are available today {\blue\citep{taira2017modal}},
we invoke the proper orthogonal decomposition (POD) given its physical underpinning (the fact that POD modes capture most energy) and its historically important role in near-wall turbulence modeling \citep{aubry1988dynamics,berkooz1993proper}.
Specifically, the mode $g$ is constructed based on a one-dimensional scalar variant of POD {\blue which is performed on the fluctuating streamwise velocity component (i.e., with the LoW subtracted) along the wall-normal direction.}

The remainder of this work is organized as follows.
In \S \ref{sec:methodology}, we provide further details of the proposed wall model including a discussion of how the $g$ mode is constructed.
In \S \ref{sec:a_priori}, we present results of an {\it a-priori} analysis of a 2D turbulent channel and a channel subjected to suddenly imposed APGs.
In \S \ref{sec:results}, {\it a-posteriori} test results are presented, again including a 2D turbulent channel and a channel with suddenly imposed APGs.
We also consider a channel with a suddenly imposed TPG.
Finally, in \S \ref{sec:conclusions}, the conclusions arrived at in this work are presented.

\section{Methodology}
\label{sec:methodology}

\subsection{Wall model formalism}
\label{subsec:formalism}

{\blue We start by considering algebraic EWMs to provide some context for our proposed extension.
In these models, we employ the LoW ansatz in \eqref{eq:EWM} as an approximation of the near-wall flow.
This is done by matching the LoW locally with the wall-parallel LES velocities such the LoW is applied along the local streamwise direction.
As equation \eqref{eq:EWM} has only one mode, i.e., the LoW mode, we can determine the coefficients ${\bf c} = (c_{x},c_{z})$ from matching information at a single off-wall location.
This gives the equations
\begin{equation}
    c_x {\rm LoW}(y^+) = U \, , \quad c_z {\rm LoW}(y^+) = W \, ,
    \label{eq:cEWM}
\end{equation}    
where $y^+$ is the wall-unit-scaled height of the matching location while $U$ and $W$ are the local wall-parallel LES velocities at this height.
We can then calculate the
wall-shear stress ${\boldsymbol \tau_w} = (\tau_{x},\tau_{z})$ from \eqref{eq:EWM} as follows
\begin{equation}
    {\boldsymbol \tau_w} = \nu \left. \frac{d{\bf U}}{dy} \right|_{y=0} = \nu {\bf c} \left. \frac{d {\rm LoW}}{dy^+} \right|_{y=0} \left. \frac{dy^+}{dy} \right|_{y=0} = {\bf c} u_\tau \, ,
    \label{eq:tau-EWM}
\end{equation}
where we have used that $dy^+/dy  = u_\tau/\nu$ and $\left.d{\rm LoW}/dy^+\right|_{y=0}=1$  by definition.
To relate the friction velocity $u_\tau$ to the coefficients ${\bf c}$, we note that EWMs assume a local state of equilibrium.
The wall-shear stress amplitude is therefore given by $\tau_w = |\boldsymbol \tau_w| = u_\tau^2$.
This, together with \eqref{eq:tau-EWM}, gives the following relation
\begin{equation}
    u_\tau = (c_x^2 + c_z^2)^{1/2} \, .
    \label{eq:utau-EWM}
\end{equation}
Thus, the appearance of the friction velocity in \eqref{eq:cEWM} (contained in $y^+$) results in this being a nonlinear set of equations that would require an iterative method to obtain the solution.
However, this requirement can be removed by using $u_\tau$ from the previous time step or from long-time averaging.
In this case, \eqref{eq:cEWM} can be solved directly and \eqref{eq:tau-EWM} is then invoked to calculate the wall-shear stress.}

{\blue The proposed extension in \eqref{eq:WM}, on the other hand}, has two modes, i.e., LoW and $g$.
To determine the coefficients ${\bf c}_{1}=(c_{1x}, c_{1z})$ and ${\bf c}_2=(c_{2x},c_{2z})$, matching with the LES at two off-wall locations is used  
\begin{equation}
\begin{split}
    &c_{1x}{\rm LoW}(y_1^+)+ c_{2x}g(y_1^+)= U_1 \, , \quad c_{1z}{\rm LoW}(y_1^+)+ c_{2z}g(y_1^+)= W_1 \, ,\\
    &c_{1x}{\rm LoW}(y_2^+)+ c_{2x}g(y_2^+)= U_2 \, , \quad c_{1z}{\rm LoW}(y_2^+)+ c_{2z}g(y_2^+)= W_2 \, .
\end{split}
\label{eq:cWM}
\end{equation}
{\blue Here, $y_1^+$ and $y_2^+$ are the wall-unit-scaled heights of the two matching locations while $U_{1,2}$ and $W_{1,2}$ are the local wall-parallel LES velocities at these heights.
Note that similar to EWMs, the augmented LoW ansatz is applied along the local streamwise direction.}
The wall-shear stress ${\boldsymbol \tau}_w = (\tau_x , \tau_z)$ can then be calculated similar to the EWM as
\begin{equation}
    {\boldsymbol \tau}_w = \nu\left. \frac{d {\bf U}}{dy}\right|_{y=0} = \nu \left( {\bf c}_1 \left. \frac{d {\rm LoW}}{dy^+} \right|_{y = 0} + {\bf c}_2 \left. \frac{d g}{dy^+} \right|_{y=0} \right) \left. \frac{dy^+}{dy} \right|_{y=0} = ({\bf c}_1 + {\bf c}_2) u_\tau \, ,
\label{eq:tau-WM}
\end{equation}
where we have again used that $dy^+/dy  = u_\tau/\nu$ and $\left.d{\rm LoW}/dy^+\right|_{y=0}=1$, while $\left. dg/dy^+\right|_{y=0} = 1$ is enforced through normalization of $g$.
{\blue Further, as for EWMs, we assume that the wall-shear stress magnitude can be expressed through a local friction velocity such that $\tau_w = | {\boldsymbol{ \tau_w}}| = u_\tau^2$ still holds for the proposed model.
Using this assumption, together with \eqref{eq:tau-WM}, we get
\begin{equation}        
    u_\tau = \left[\left(c_{1x}+c_{2x}\right)^2+\left(c_{1z}+c_{2z}\right)^2\right]^{1/2} \, .
    \label{eq:utau-WM}
\end{equation}
Comparing with the EWM result from \eqref{eq:utau-EWM}, we see that the structure is very similar.
Further, the friction velocity $u_\tau$ in \eqref{eq:cWM} (contained in $y_1^+,y_2^+$) results in a set of nonlinear equations, which can be handled in the same way as for the EWM.
With the coefficients determined, \eqref{eq:tau-WM} is then invoked to compute the wall-shear stress.}
It is worth noting that the model reduces to an algebraic EWM when ${\bf c}_2=0$.

Following \cite{yang2017log}, \blue we apply additional filtration to the velocities at both matching locations.
This filtration allows us to place the matching location(s) close to the wall without a log-layer mismatch, which is preferred to placing the matching location(s) further away; since the latter incurs penalties on parallel computing, among other disadvantages.
\blue \cite{yang2017log} also found similar behaviors using temporal and spatial filters in 2D channel flow cases.
We have chosen to only consider spatial filtering in this work since the model will be validated against highly unsteady flows, for which spatial filtering is more appropriate.
\black

\subsection{Wall model modes}
\label{subsec:POD}

We now describe the two modes, i.e., ${\rm LoW}$ and $g$, in greater detail.
The construction of the $g$ mode is covered separately in \S \ref{subsec:g_mode_const} below, {\bluenew and we discuss the effect of the choice of the $g$ mode in Appendix \ref{App:choice_of_g_mode}}. 
Figure \ref{fig:modes}(a) shows the LoW mode.
This is straightforward.
The mode contains the viscous sublayer, the buffer layer, and the logarithmic layer, but not the wake layer.
The behavior of the LoW mode conforms to $U^+=\log(y^+)/\kappa+B$ at sufficiently large $y^+$.
\blue Analytical expressions for the viscous sublayer and buffer layer can be found in \cite{reichardt1951vollstandige} and \cite{spalding1961single}\black.  
Figure \ref{fig:modes}(b) shows the POD-based mode $g$. 
The mode is approximately constant away from the wall and conforms to the no-slip condition at the wall.

\begin{figure}
\centering
\includegraphics[width=0.32\textwidth]{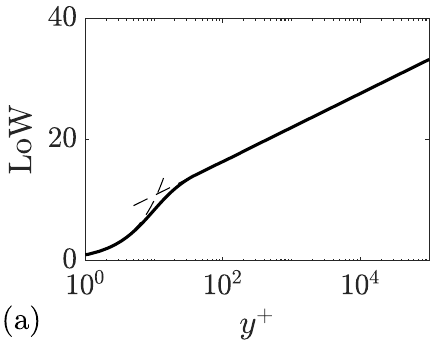}~~~~\includegraphics[width=0.32\textwidth]{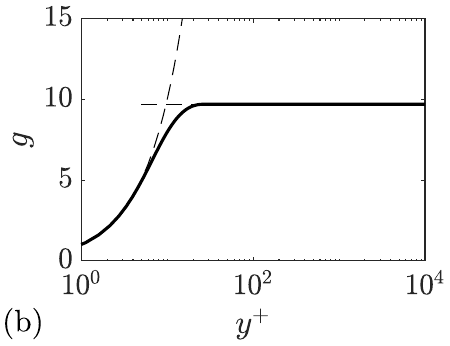}
\caption{\label{fig:modes} (a) The LoW mode. The two dashed lines correspond to ${\rm LoW} = y^+$ and ${\rm LoW} = \log(y^+)/\kappa+B$. 
(b) The $g$ mode. 
The two dashed lines correspond to $g=y^+$ and $g=9.7$.
} 
\end{figure}

{\blue A basic requirement of a wall model is that it degenerates to the no-slip condition at the wall-resolved limit.
Here, We discuss the behavior of the model when the viscous sublayer is resolved.
When the viscous layer is resolved, we have 
\begin{equation}
{\bf U}=\left.\frac{d {\bf U}}{dy}\right|_{y=0} y \, ,
\label{eq:visc_sub_approx}
\end{equation}
at first off-wall grid points.
As seen in figure \ref{fig:modes}, both the LoW and the mode $g$ are linear functions of $y^+$ in the viscous sublayer, so in this case, we have
\begin{equation}
{\rm LoW} = g = y^+ \, .
\label{eq:LoW_and_g_in_visc_sub}
\end{equation}
It follows from \eqref{eq:visc_sub_approx} and \eqref{eq:LoW_and_g_in_visc_sub} that \eqref{eq:cWM} reduces to 
\begin{equation}
  ({ {\bf c}_{1}+{\bf c}_{2}})y_1^+=\left.\frac{d{\bf U}}{dy}\right|_{y=0} y_1 \, , \quad
  ({ {\bf c}_{1}+{\bf c}_{2}})y_2^+=\left.\frac{d{\bf U}}{dy}\right|_{y=0} y_2 \, .
\end{equation}
The two equations have the following solution
\begin{equation}
  { {\bf c}_{1}+{\bf c}_{2}}=\frac{\bf U}{|{\bf U}|}u_\tau \, .
  \label{eq:c1c2}
\end{equation}
Although ${\bf c}_1$ and ${\bf c}_2$ are not uniquely determined, ${ {\bf c}_{1}+{\bf c}_{2}}$ is.
As a result, the wall-shear stress is uniquely determined.
We also notice that \eqref{eq:c1c2} is non-singular even when the wall-shear stress is 0.
Hence, $\tau_w=0$ is a removable singularity
($\tau_w=0$ is a singularity because $y^+$ is not defined when $\tau_w=0$, it is removable because it does not create any mathematical singularity to the calculation of the wall-shear stress).
In terms of numerical implementation, one could choose to do a minimum norm solution, which would be quite elegant.
We have chosen to add a simple ``if'' statement that allows the code to treat wall-resolved grids separately.}

Finally, when implementing the wall model, we tabulate the LoW mode and the $g$ mode in a lookup table.
An inquiry about LoW's and $g$'s values at a given $y^+$ is then interpolated from the lookup table.
{\blue Here, we prefer lookup tables over analytical formulation because ``addressing'' a lookup table is more efficient than computing a closed-form expression.
In computer science terms, a lookup table is an array that replaces runtime computation with a simpler array indexing operation. 
The tables may be precalculated and stored in static program storage, pre-fetched as part of a program's initialization, or stored in hardware in application-specific platforms. 
The process is called ``direct addressing''.
Because retrieving a value from memory is often faster than carrying out a more expensive computation or input/output operation, lookup tables can save processing time.}

\subsection{Construction of the \textbf{\textit{g}} mode}
\label{subsec:g_mode_const}

Define ${\bf r} = {\bf U}-{\bf c}_1 {\rm LoW}$ such that ${\bf r} = (r_x,r_z)$ describes the deviation from the LoW.
The $g$ mode, or more precisely ${\bf c}_2 g$, is intended to capture ${\bf r}$.
Any function, as long as it is not exactly the LoW, should capture part of that deviation.
Here, we want a mode that captures as much energy in $\bf r$ as possible.
This directly leads to POD.
Specifically, we consider a one-dimensional scalar variant of POD.
It is POD of the \blue fluctuating \black streamwise velocity component along the wall-normal direction in a 2D turbulent channel.
Further details on the numerical implementation are given in Appendix \ref{App:Num_solv_POD_eig_prob}.
{\blue As POD can be performed in multiple spatial dimensions, we provide a few observations in this regard.
First, if the POD analysis were to be performed in 3D space, a local coupling between the wall model and LES would no longer be possible.
Instead, a more global coupling over some wall-parallel plan(s) would be necessary which could be limiting for flows in complex geometries.
Second, a global coupling between LES and the wall model could be problematic in the parallelization of the WMLES code.}

\begin{figure}
\centering
\includegraphics[width=0.8\textwidth]{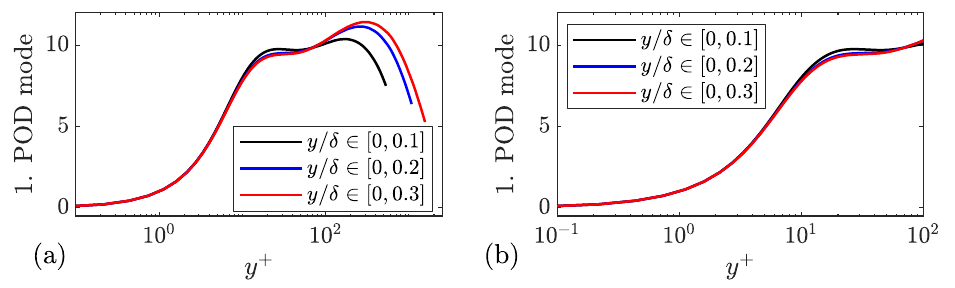}
\caption{\label{fig:g_mode_const} (a) First POD mode for three wall-normal intervals which start from the wall and end at $y/\delta = 0.1$, $0.2$, and $0.3$. 
(b) Zoom in on the near-wall region.
} 
\end{figure}

\begin{figure}
\centering
\includegraphics[width=0.8\textwidth]{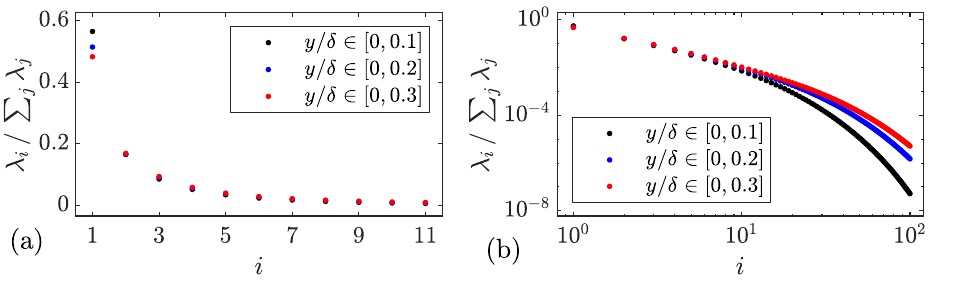}
\caption{\label{fig:POD_spectra} (a) POD eigenvalue spectra for the three cases in figure \ref{fig:g_mode_const}. 
(b) Same as in (a) but in a double logarithmic plot.
} 
\end{figure}

We perform this one-dimensional POD for wall-normal intervals ranging from the wall to the log-layer region in a 2D turbulent channel with $Re_\tau=5200$ {\blue \citep{lee2015direct}}.
Three different heights within the log-layer are considered.
The result for the first POD mode is shown in figure \ref{fig:g_mode_const}. 
We see that the near-wall part of the POD mode does not change and that the modes asymptote to a constant in the logarithmic layer.
This gives rise to the $g$ mode in figure \ref{fig:modes}(b).
{\blue Further, the POD spectra for the three different analyses are given in figure \ref{fig:POD_spectra}.
In figure \ref{fig:POD_spectra}(a), it can be observed that the first and second POD modes carry roughly 50\% and 15\%  of the total energy, respectively, for all three cases.
This shows that the first POD mode is significantly more important than the subsequent modes.}
It is curious to note that the $g$ mode used here is almost identical to the laminar mode in {\blue\cite{fowler2022lagrangian}}, where the mode describes the response of the flow to an alternating pressure gradient.
{\blue This suggests that the instantaneous flow in an equilibrium channel is also subjected to large instantaneous pressure gradients.}

\begin{figure}
\centering
\includegraphics[width=0.4\textwidth]{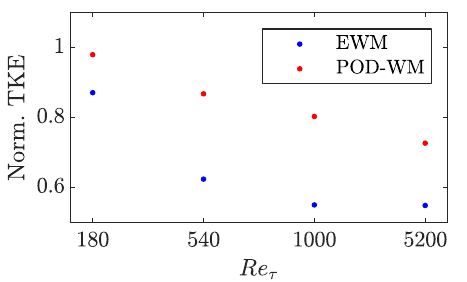}
\caption{\label{fig:TKE_WM_test} Normalized streamwise TKE captured by the EWM in \eqref{eq:EWM} and POD-augmented WM in \eqref{eq:WM} for different $Re_\tau$.
The normalization is by the TKE from DNS.} 
\end{figure}

\blue We also comment on the choice of $Re_\tau =$5200 for the POD analysis.
As WMLES is primarily aimed at high Reynolds number flows, it is preferable to base the POD analysis on such a case.
Currently, $Re_\tau = 5200$ is the largest Reynolds number for which the instantaneous velocity field, needed for the POD analysis, is publicly available \citep{graham2016web}.
Still, an ideal wall model should be able to perform well also at low Reynolds numbers.
Therefore, we have performed a simple test at different $Re_\tau$ for both the EWM \eqref{eq:EWM} and POD-augmented WM \eqref{eq:WM}, with the $g$ mode from $Re_\tau = 5200$ in figure \ref{fig:modes}(b), to compare them in terms of capturing the turbulent kinetic energy (TKE).
We do this using DNS data from four different Reynolds numbers: $Re_\tau = 180$, 540, 1000, and 5200.
The data for the two cases $Re_\tau = 180$ and $540$ are generated using a code with the same numerics as in \cite{Kim_et_al_1987_DNS_channel} while the cases $Re_\tau = 1000$ and 5200 are from \cite{lee2015direct,graham2016web}.
We then fit the streamwise coefficients of both wall models in \eqref{eq:EWM} and \eqref{eq:WM} by a least squares regression using the streamwise velocity from DNS data over the interval $0 \leq y / \delta \leq 0.1$.
This interval is chosen to be consistent with the subsequent {\it a-priori} and {\it a-posteriori} testing for a 2D turbulent channel in \S \ref{sec:a_priori} and \S \ref{sec:results}, respectively.
The results are shown in figure \ref{fig:TKE_WM_test}.
We observe that the POD-augmented WM captures significantly more of the TKE than the EWM at all Reynolds numbers considered.
In regards to the decrease in the captured TKE with increasing Reynolds number, which is observed for both wall models, this can be explained by analogy with the momentum cascade in homogeneous isotropic turbulence.
Specifically, as the Reynolds number increases, smaller and smaller scales are activated in the flow which then start contributing to the total TKE.
As these small scales cannot be captured by a few large-scale modes, the relative fraction of the TKE captured by such a finite set of large-scale modes will inevitably decrease as the Reynolds number increases\black.

Admittedly, the POD modes in another flow will not be the same as the ones in a 2D channel, {\blue but we argue that this is not a huge concern here given the purposes of this work.
Specifically, considering POD modes from other wall-bounded turbulent flows will only be meaningful if the present approach yields more accurate results than the EWM.
If this turns out to be the case, a more detailed analysis involving POD modes from other flows should be considered to determine the potential benefits of using particular modes in different flow types.
Still, we note that the flow physics that governs the instantaneous flow in a channel is also present in other wall-bounded flows.
Therefore, wall modeling improvements based on a mode from this case would be an encouraging sign for application in other flow types as well.}
{\blue Last, we comment on the choice of the flow configuration.
Considering that the inner layer of a boundary layer is no different from that of a channel, POD analysis of a boundary-layer flow should give us similar results.
Nonetheless, DNS data of channel flows is more extensively available than DNS data of boundary layers.}

\section{A-priori analysis}
\label{sec:a_priori}

We perform a comparative {\it a-priori} analysis between the EWM and the POD-augmented WM.
We consider a 2D equilibrium channel and channels subjected to a strong adverse pressure gradient.
We will show that the present POD-augmented WM captures non-equilibrium effects more accurately than the EWM.

\subsection{Equilibrium channel}

\begin{figure}
\centering
\includegraphics[width=0.8\textwidth]{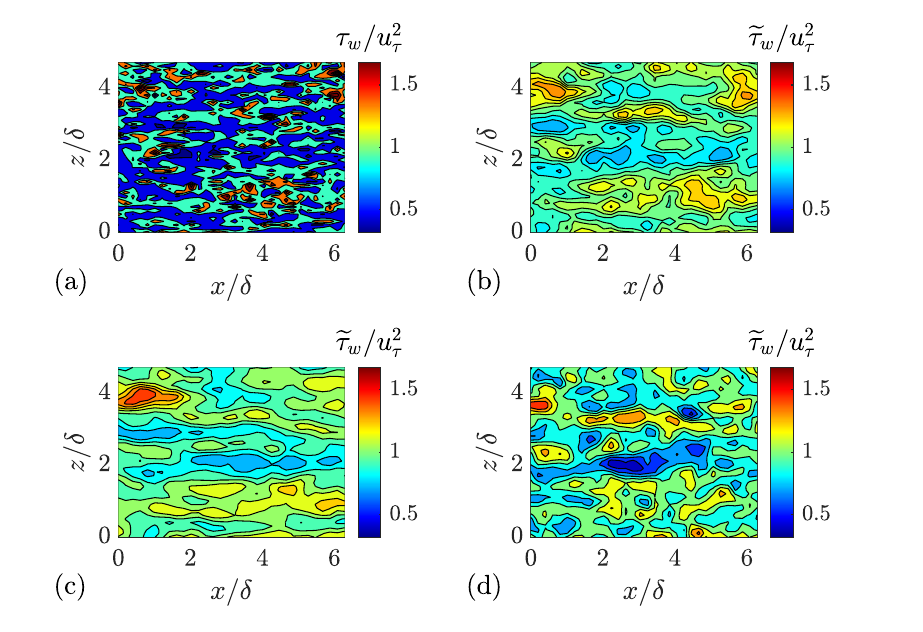}
\caption{\label{fig:WM_tauw_a_priori_channel} (a) Wall-shear stress from DNS. (b) Wall-shear stress from filtered DNS.\\(c) Wall-shear stress predicted by the EWM. (d) Wall-shear stress predicted by the POD-augmented WM. The WM calculations are discussed in the text.
Note that we show only part of the computational domain for presentation purposes.
}
\end{figure}

\begin{figure}
\centering
\includegraphics[width=1\textwidth]{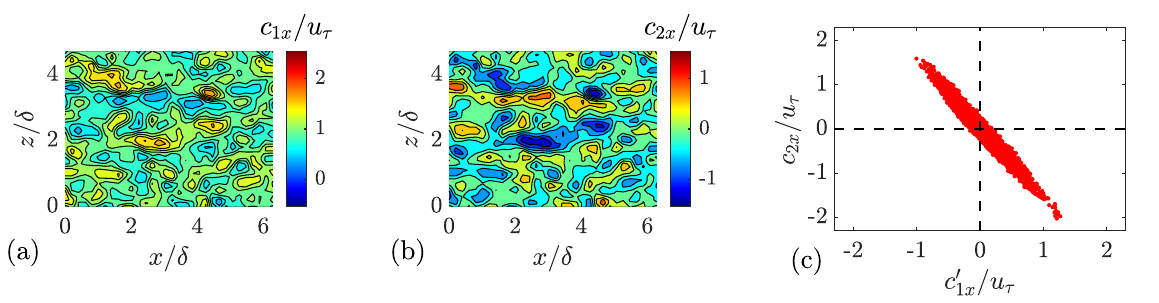}
\caption{\label{fig:WM_coeffs_a_priori_channel} (a) Contours of $c_{1x}$. (b) Contours of $c_{2x}$. (c) Instantaneous realizations of $c_{1x}'$ and $c_{2x}$. 
} 
\end{figure}

We consider a 2D turbulent channel with $Re_\tau = 1000$ \citep{graham2016web}.
We filter the DNS data following \cite{graham2016web} and \cite{yang2019predictive} so that the study is more faithful to WMLES than DNS.
A Gaussian filter is used with a filtration length scale of $300 (\nu / u_\tau)$.
{\blue This length scale is chosen to match the grid resolution used in the subsequent {\it a-posteriori} investigations in \S \ref{subsec:channel} to make the results more comparable.}
The filtered DNS data is then used to fit for {\blue the coefficients in \eqref{eq:EWM} and \eqref{eq:WM}.
Note that we use the same DNS data for fitting both the EWM and POD-WM to ensure a fair comparison.}
{\blue Specifically, the matching locations are at $y/\delta=0.5/12$ and 1.5/12, i.e., the first and the second off-wall grid point locations in the subsequent {\it a-posteriori} study in \S \ref{subsec:channel}}.
The wall-shear stress is computed according to \blue \eqref{eq:tau-EWM} for the EWM and \eqref{eq:tau-WM} for the POD-augmented WM\black.
The wall-shear stress computed from the EWM and POD-augmented WM is compared with the wall-shear stress from both DNS and filtered DNS in figure \ref{fig:WM_tauw_a_priori_channel}.
We observe the following.
First, both the EWM and POD-augmented WM predict the correct mean stress.
Second, the EWM captures the large-scale variations of the wall-shear stress, but it does not capture the more extreme wall-shear stress events.
On the other hand, the POD-augmented WM captures the large-scale variation and some of the extreme events.
\blue For a more quantitative comparison, we note that \black the root-mean-square (rms) of the wall-shear stress fluctuations $\tau_{w,{\rm rms}}^+$ is 0.42, 0.13, 0.11, and 0.20 in DNS, filtered DNS, predicted by the EWM, and predicted by the POD-augmented WM, respectively.
\blue Further, we also give the rms of the error between the filtered DNS and the EWM and POD-WM predictions which are 0.09 and 0.16, respectively.
Thus, we see that the rms of the error is larger for the POD-augmented WM than the EWM when the models are compared with filtered DNS data\black.
As the POD-augmented WM contains two modes, i.e., the LoW and POD-based mode $g$, it is interesting to investigate the behavior of the coefficients in front of these two modes.
Figure \ref{fig:WM_coeffs_a_priori_channel} shows the contours of these coefficients and their correlation.
First, we observe that the two coefficients are of the same magnitude.
Second, there is a strong anti-correlation between the LoW mode and the $g$ mode coefficients.
This suggests that when the equilibrium LoW term deviates from the mean value, the non-equilibrium term pulls it back.
{\blue We have repeated the above {\it a-priori} test at other Reynolds numbers and matching locations.
We see that $c_{1x}$ and $c_{2x}$ are always anti-correlated, but the correlation depends on the Reynolds number and the matching location.}
The exact physical mechanism behind this observation is not entirely clear to us, but this phenomenon, if it translates to WMLES, will help to stabilize the numerical simulation.

\subsection{Non-equilibrium channel}
\label{subsec:non-equilib-a-priori}

The main purpose of the POD-augmented WM is to capture non-equilibrium effects that are not captured by the EWM.
In this subsection, we consider the time-evolution of a 2D channel flow subjected to suddenly applied APGs. 
Using the nomenclature {\blue R[$Re_\tau$]A[APG$/(\tau_{w,0}/\delta)$]}, we consider the two cases: {\blue R1000A10 and R1000A100} with moderate and strong adverse pressure gradients.
Again, we fit for the coefficients in \eqref{eq:EWM} and \eqref{eq:WM} {\blue using the same matching information for both models with the matching locations being the same as those used in the {\it a-posteriori} study in \S \ref{subsec:APG}. 
The wall-shear stress is computed according to \eqref{eq:tau-EWM} and \eqref{eq:tau-WM}.}
We then compare the computed wall-shear stresses to the DNS.

\begin{figure}
\centering
\includegraphics[width=0.8\textwidth]{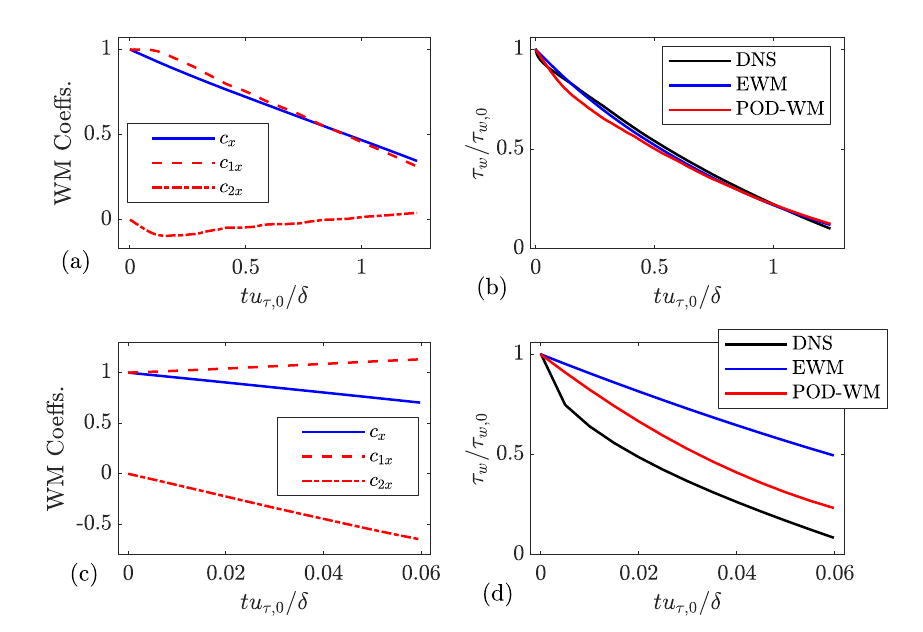}
\caption{\label{fig:WM_a_priori_APG} (a) WM coefficients {\blue $c_x$ from the EWM and $c_{1x}$ and $c_{2x}$ from the POD-augmented WM} in the case {\blue R1000A10}. 
(b) Wall-shear stress {\blue from DNS, the EWM, and the POD-augmented WM} in case {\blue R1000A10}.
(c, d) Same as (a, b) but for case {\blue R1000A100}.
}
\end{figure}

Figure \ref{fig:WM_a_priori_APG}(a, b) shows the {\blue R1000A10} results.
We see from figure \ref{fig:WM_a_priori_APG}(a) that there is a good agreement between the LoW coefficients $c_x$ and $c_{1x}$ from the EWM and POD-augmented WM, respectively, and that the $g$ mode in the POD-augmented WM is almost entirely ``turned off'' as $c_{2x} \approx 0$.
Hence, the POD-augmented WM formulation reduces to the EWM in near-equilibrium conditions.
In regards to the accuracy of the wall-shear stress predictions, we see that both the EWM and POD-augmented WM produce accurate results as seen in figure \ref{fig:WM_a_priori_APG}(b).
\blue Still, it should be observed that the POD-augmented WM performs slightly worse than the EWM for this near-equilibrium case\black. 
Next, the {\blue R1000A100} results are shown in figure \ref{fig:WM_a_priori_APG}(c, d).
From figure \ref{fig:WM_a_priori_APG}(c), we see that the LoW coefficients for the EWM and POD-augmented WM no longer agree and that the $g$ mode in the POD-augmented WM is active.
Specifically, for the POD-augmented WM, we observe that the LoW coefficient $c_{1x}$ remains almost constant while the $g$ coefficient $c_{2x}$ continuously decreases.
Further, the wall-shear stress predictions in figure \ref{fig:WM_a_priori_APG}(d) show that the POD-augmented WM does a much better job of capturing the strong decrease in the wall-shear stress than the EWM.

\section{A-posteriori studies}
\label{sec:results}

Good results in {\it a-priori} analysis do not necessarily translate to good results in {\it a-posteriori} studies.
We apply the POD-augmented WM to an equilibrium 2D channel, and channels subjected to a suddenly imposed adverse or transverse pressure gradient.

\subsection{Code}
\label{subsec:code}

The POD-augmented WM is implemented in our in-house pseudo-spectral code LESGO.
The code solves the incompressible Navier-Stokes equations in a half-channel with a slip top boundary and periodicity in both the streamwise and the spanwise directions.
The flow is driven by a constant streamwise pressure gradient.
The code employs a pseudo-spectral method in the streamwise and the transverse directions, a second-order finite difference method in the wall-normal direction, and the second-order Adam-Bashforth method for time stepping.
The grid is uniform in each direction.
The time step size is such that the CFL number is always smaller than 0.06 (\blue due to the Adams-Bashforth time-stepping scheme\black).
{\blue Further, a Gaussian filter is applied to the wall-parallel velocities before they are used as input to the wall model.}
Last, the sub-grid scale (SGS) stresses are modeled via the Lagrangian scale-dependent model \citep{bou2005scale}.
The code is well-validated \citep{abkar2017large,yang2022logarithmic} and further details of the code's numerics can be found in \cite{yang2018hierarchical} and the references cited therein. 

\subsection{Fully developed channel}
\label{subsec:channel}

\begin{figure}
\centering
\includegraphics[width=0.8\textwidth]{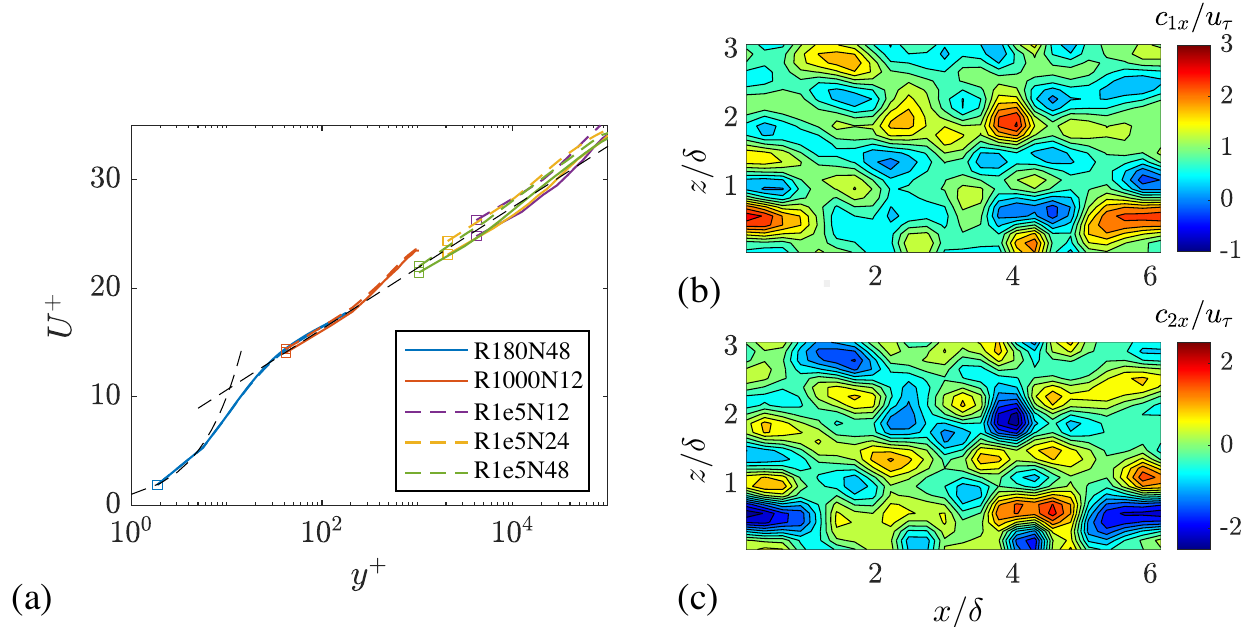}
\caption{\label{fig:chan} (a) Mean velocity profiles. The square symbols indicate the location of the first off-wall grid point.
The solid lines are POD-WM results, and the dashed lines are EWM results.
(b) Contours of $c_{1x}$. (c) Contours of $c_{2x}$.
} 
\end{figure}

We first apply the POD-augmented WM in 2D equilibrium channels.
This exercise will serve as validation: the WM must be able to predict the law of the wall at all Reynolds numbers on both wall-resolved and wall-modeled grids.
Table \ref{tab:chan} shows the details of the WMLES.
The nomenclature of the cases is {\blue R[$Re_\tau$]N[$N_y$]}. 
The Reynolds number is from $Re_\tau=180$ to $10^5$, the grid is from DNS-like (in case {\blue R180N48}) to typical-WMLES-like ({\blue R1000N12, R1e5N12}), {\blue and the domain is of the size as in \cite{lozano2014effect}}.
The LES/WM matching locations are at the first two off-wall grid points, {\blue which are used by both wall models.}
The matching locations are in the viscous sublayer (in case {\blue R180N48}) or the logarithmic layer (in cases {\blue R1000N12, R1e5N12, R1e5N24, R1e5N48}).
Following the established best practice \citep{larsson2015large}, the matching locations are not placed in the buffer layer.
{\blue Placing the matching location within the buffer layer incurs log-layer mismatch for WMLES in general---even if the matching location is not the first off-wall grid point and the velocity at the matching location is filtered.
Nonetheless, placing the matching location is rarely a concern for flows at high Reynolds numbers since the buffer layer will not be resolved.
This issue comes up here only because we validate against DNS data, which is limited to moderate Reynolds numbers.}
Figure \ref{fig:chan}(a) shows the mean flow.
The profiles follow the LoW irrespective of the grid and the Reynolds number.
{\blue We see some difference at high Reynolds numbers when the grid resolution is low, which may be due to the interplay between the SGS model and the wall model.}
Figure \ref{fig:chan}(b, c) shows the instantaneous contours of $c_{1x}$ and $c_{2x}$.
The mean flow is $U=u_\tau~{\rm LoW}(y^+)$ in a channel.
Hence, $c_{1x}$'s mean should be $u_\tau$ and $c_{2x}$'s mean should be $0$.
The above expectation bears out in figure \ref{fig:chan}(b, c) {\blue where $c_{1x} / u_\tau = 0.96$ and $c_{2x} / u_\tau = 0.05$ when averaged over the wall-parallel directions.
We note that the observed deviation from $c_{1x} = u_\tau$ and $c_{2x} = 0$ is due to instantaneous effects as they disappear upon time averaging.}
Further, the local instantaneous deviations in $c_{1x}$ from $u_\tau$ is a result of non-equilibrium effects, where we should see the POD-based $g$ mode.
This expectation also bears out in figure \ref{fig:chan}(b, c).

\begin{table}%Table of simulation parameters
\begin{center}\scriptsize
\vskip -0.1in
\begin{tabular}{cccccc}
~~~~Case~~~~   & ~~~~$Re_\tau$~~~~ & ~~~~Grid~~~~                   & ~~~~Domain~~~~                  & ~~~~ {\blue$\Delta y^+/2$}~~~~ & ~~~~ WM ~~~~\\ 
{\blue R180N48}  & 180       & $96\times 48\times 96$ & $2\pi\times 2\times \pi$ & 1.88  & EWM, POD-augmented WM         \\
{\blue R1000N12} & 1000      & $24\times 12\times 24$ & $2\pi\times 2\times \pi$ & 41.7    & EWM, POD-augmented WM       \\ 
{\blue R1e5N12} & $10^5$      & $24\times 12\times 24$ & $2\pi\times 2\times \pi$ & 4.17$\times 10^3$ & EWM, POD-augmented WM  \\
{\blue R1e5N24} & $10^5$          & $48\times 24\times 48$ & $2\pi\times 2\times \pi$ & $2.08\times 10^3$ & EWM, POD-augmented WM\\ 
{\blue R1e5N48} & $10^5$           & $96\times 48\times 96$ & $2\pi\times 2\times \pi$ & $1.04\times 10^3$ & EWM, POD-augmented WM\\ 
\end{tabular}
\caption{\label{tab:chan}WMLES details. 
The half-channel height is used to normalize the numbers in the ``Domain'' column. 
The first off-wall grid point is at $\Delta y/2$ (the half is because of the staggered grid).}
\end{center}
\end{table}

\subsection{Adverse pressure gradient}
\label{subsec:APG}

Next, we apply the POD-augmented WM in a channel subjected to a suddenly imposed APG.
A schematic of the flow was shown in figure \ref{fig:sketch-chan}(a) already, and
figure \ref{fig:sketch-chan}(c) shows the time evolution of the mean flow.
Table \ref{tab:APG} shows the WMLES details. 
The nomenclature is {\blue R[$Re_\tau$]A[APG$/(\tau_{w,0}/\delta)$]}, where the subscript 0 indicates a variable evaluated prior to the application of the APG.
The channel is initially at a Reynolds number $Re_\tau=1000$.
The APGs are 10$(\tau_{w,0}/\delta)$ and 100$(\tau_{w,0}/\delta)$ in cases {\blue R1000A10 and R1000A100}, respectively.
The flow decelerates, and the Reynolds number decreases.
LES information at the second and the fourth off-wall grid points are fed to the POD-augmented WM so that we do not match in the buffer layer.
For a fair comparison, the EWM makes use of the LES information at these two off-wall locations as well.
Specifically, the EWM predicts wall-shear stress that yields the best fit of the velocity through the second and fourth off-wall grid points.

\begin{table}%Table of simulation parameters
\begin{center}\scriptsize
\vskip -0.1in
\begin{tabular}{cccccc}
~~~~Case~~~~    & ~~~~$Re_{\tau,0}$~~~~ & ~~~~Grid~~~~                   & ~~~~Domain~~~~                   & ~~~~APG~~~~ & ~~~~ WM ~~~~ \\ 
{\blue R1000A10N12}  & 1000          & $24\times 12\times 24$ & $2\pi\times 2\times \pi$ & 10 & EWM, POD-augmented WM \\
{\blue R1000A100N12} & 1000          & $24\times 12\times 24$ & $2\pi\times 2\times \pi$ & 100 & EWM, POD-augmented WM\\
\end{tabular}
\caption{\label{tab:APG}WMLES details. 
The half-channel height is used to normalize the numbers in the ``Domain'' column. 
The APG column shows the magnitude of the suddenly imposed APG.
The numbers are normalized using $\tau_{w,0}/\delta$.
}
\end{center}
\end{table}

\begin{figure}
\centering
\includegraphics[width=0.8\textwidth]{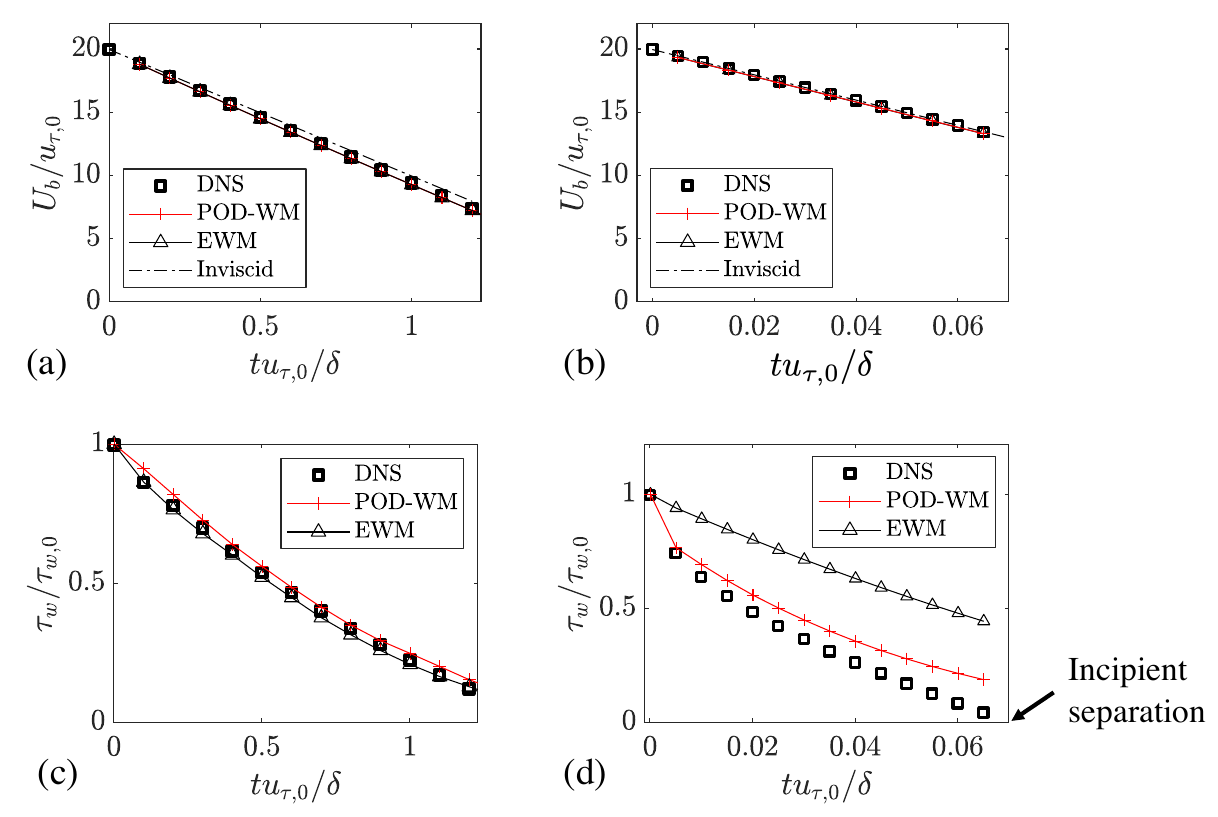}
\caption{\label{fig:APG1} (a, b) Flow rate as a function of time. (a) {\blue R1000A10}. (b) {\blue R1000A100}. Here, ``Inviscid'' corresponds to   \eqref{eq:ub}. 
(c, d) Wall-shear stress as a function of time. (c) {\blue R1000A10}. (d) {\blue R1000A100}.
DNS: values given by the DNS, which we take as the truth.
POD-WM: predictions by the WMLES that employs the POD-augmented WM for near-wall turbulence modeling.
EWM: predictions by the WMLES that employs the EWM for near-wall turbulence modeling.
} 
\end{figure}

Figure \ref{fig:APG1}(a, b) shows the flow rate as a function of time for {\blue R1000A10 and R1000A100}, respectively, and we compare WMLES and DNS.
The DNS data are by the present authors as well.
The setups are similar to those in \cite{he2015transition} and are not detailed here for brevity.
We see from figure \ref{fig:APG1}(a, b) that the flow rate is accurately predicted irrespective of the wall model.

We explain why wall modeling is not critical to the prediction of the flow rate and the velocity itself.
The flow rate is governed by the following volume-integrated $x$ momentum equation, ${d(\rho U_b)}/{dt}=-{\tau_w}/{\delta}-{dP}/{dx}$.
If the flow is inviscid, this equation becomes ${d(\rho U_b)}/{dt}=-{dP}/{dx}$, which directly leads to
\begin{equation}
    U_b=U_{b,0}-\frac{1}{\rho}\frac{dP}{dx}t \, . 
    \label{eq:ub}
\end{equation}
The above inviscid estimate is plotted in figure \ref{fig:APG1}(a, b) and is fairly accurate.
{\bluenew In fact, the imposed adverse pressure gradient overwhelms the wall-shear stress and dominates the evolution of the velocity field.
This makes wall modeling less critical to the prediction of the flow rate.
}

A more difficult test is the wall-shear stress.
Figure \ref{fig:APG1}(c, d) shows the (horizontally averaged) wall-shear stress as a function of time for {\blue R1000A10 and R1000A100}, respectively.
The EWM captures the wall-shear stress in case {\blue R1000A10} but grossly over-predicts the wall-shear stress in case {\blue R1000A100}.
The POD-augmented WM, on the other hand, captures the wall-shear stress well in both cases---although its prediction becomes less accurate as the flow approaches separation in case {\blue R1000A100}.
{\blue We also observe that the POD-augmented WM is slightly less accurate than the EWM for the R1000A10 case which was also observed in the {\it a-priori} tests.}

\begin{figure}
\centering
\includegraphics[width=1\textwidth]{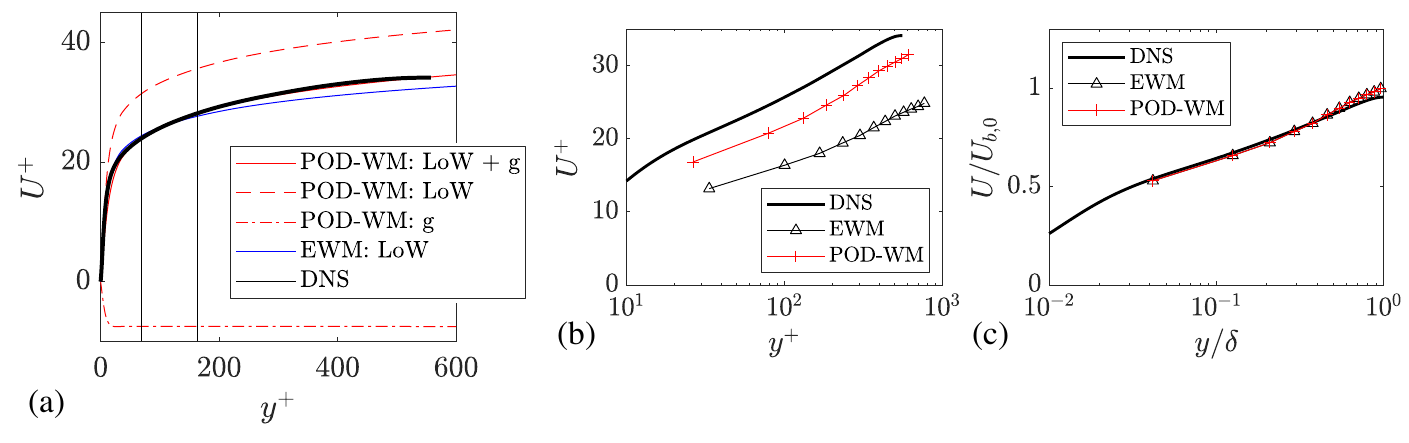}
\caption{\label{fig:APG2} (a) DNS profiles and {\it a-priori} WM reconstructions at $tu_{\tau,0}/\delta=0.035$ for case {\blue R1000A100}. 
The two vertical lines are at $1.5/12\delta$ and $3.5/12\delta$, which correspond to the 2nd and 4th grid points in our WMLES.
The LoW mode and the $g$ mode in the POD-augmented WM are also plotted.
{\blue This corresponds to $c_{1x}$LoW and $c_{2x}g$.}
Their contributions to the wall-shear stress are of opposite signs, which is consistent with figure \ref{fig:chan}.
(b, c) Mean velocity profiles from DNS, WMLES with EWM and POD-augmented WM,
(b) normalization by inner units,
(c) normalization by outer units.
} 
\end{figure}

{\black We have shown in \S \ref{sec:a_priori} that the POD-augmented wall model captures non-equilibrium effects.
Here, we explain why the POD-augmented WM is more accurate than the EWM.
{\bluenew We note that the velocity and the wall-shear stress are coupled, and the accurate prediction of one depends on the accurate prediction of the other.
Nonetheless, for the two problems here, the imposed pressure gradient overwhelms the wall-shear stress and dominates the evolution of the velocity.
Consequently, the task of wall modeling here simplifies to the prediction of the wall-shear stress when given the velocity.}
Under this simplification, a WM is accurate if it gives an accurate reconstruction of the near-wall velocity field.
Figure \ref{fig:APG2}(a) shows the DNS velocity profile at $tu_{\tau,0}/\delta=0.035$ in case R1000A100 as well as the reconstructions of the mean velocity profiles according to the POD-augmented WM and the EWM.
We see from figure \ref{fig:APG2}(a) that the POD-augmented WM provides a more accurate reconstruction of the mean flow than the EWM.}

{\bluenew
We can then quantify the error in the predicted wall-shear stress by computing the wall-shear stress due to the LoW mode and the $g$ mode.
Table \ref{tab:APG-tauw} tabulates results.
The LoW mode in the EWM gives $\tau_w/\rho=0.50 u_{\tau,0}^2$, leading to a 61\% error.
The LoW mode and the $g$ mode in the POD-WM give $0.62 u_{\tau,0}^2$ and $-0.24 u_{\tau,0}^2$, leading to an overall error of 23\%.
Table \ref{tab:APG-tauw} also tabulates the error in figure \ref{fig:APG1} at the same time instant.
There, the errors are 116\% and 29\%.
The larger error in the {\it a-posteriori} test is because the error has accumulated in the WMLES.}

\begin{table}%Table of simulation parameters
\begin{center}
\vskip -0.1in
\begin{tabular}{ccccccc}
\multirow{2}{*}{} & \multicolumn{4}{c}{\it Reconstruction} & \multicolumn{2}{c}{{\it WMLES}} \\
                  & ~~$\tau_w$: LoW~~   & ~~$\tau_w$: g~~  & ~~$\tau_w$~~  & ~~error~~  & ~~$\tau_w$~~                     & ~~error~~                    \\
EWM               & 0.50            & 0            & 0.50      & 61\%   & 0.67                         & 116\%                    \\
POD-augmented WM            & 0.62            & -0.24        & 0.38      & 23\%   & 0.40                         & 29\%                    
\end{tabular}
\caption{\label{tab:APG-tauw}Contributions of each mode to the wall-shear stress at time $tu_{\tau,0}/\delta=0.035$ in case {\blue R1000A100}.
The EWM does not contain a $g$ mode, and therefore its contribution to EWM's prediction is 0.
The reconstruction result corresponds to figure \ref{fig:APG2}, and the WMLES result corresponds to figure \ref{fig:APG1}. 
The wall-shear stress values are normalized with $\rho u^2_{\tau,0}$.
The DNS value is 0.31, which we take as the truth.
}
\end{center}
\end{table}

Before we conclude this section, we comment on the presentation of the results.
Figure \ref{fig:APG2}(b, c) show the DNS and WMLES profiles at $tu_{\tau,0}/\delta=0.035$ in case {\blue R1000A100}.
At this time instant, the DNS predicts $\tau_w=0.31\tau_{w,0}$. 
The POD-augmented WM predicts $\tau_w=0.40\tau_{w,0}$, and the EWM predicts $\tau_w=0.67\tau_{w,0}$, leading to a 116\% error in EWM's prediction and a 29\% error in POD-augmented WM's prediction.
The EWM is not at all accurate.
This, however, is only clear in figure \ref{fig:APG2}(b), where normalization is by the inner units, not in (c), where normalization is by the outer units.
{\bluenew Again, this is because the imposed adverse pressure gradient overwhelms the wall-shear stress, and the evolution of the velocity profile is dominated by the imposed pressure gradient for the initial period.}
{\blue We would also like to comment on the flow statistics we show. 
WMLES and LES, in general, resolve only part of the energy spectrum, and therefore they are not meant to capture turbulence quantities like the velocity rms and the energy spectrum, which is why we have chosen to focus on mean flow quantities.}

\subsection{Transverse pressure gradient}
\label{subsec:TPG}

\begin{figure}
\centering
\includegraphics[width=0.8\textwidth]{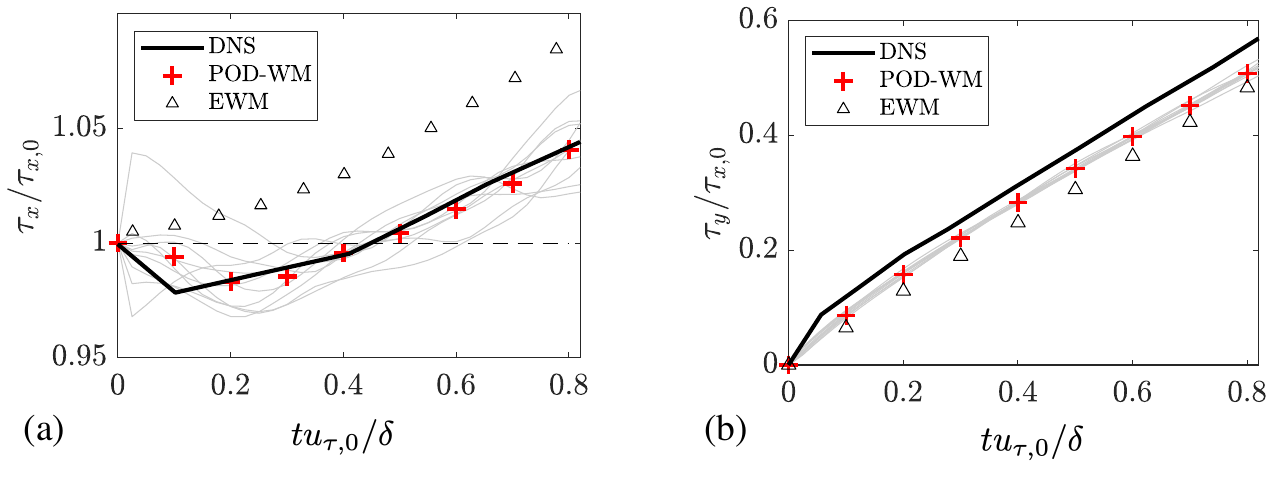}
\caption{\label{fig:SPG1} Time evolution of (a) the streamwise wall-shear stress and (b) the spanwise wall-shear stress.
The DNS is by \cite{lozano2020non}.
The POD-augmented WM results are ensemble averages from 10 statistically independent realizations.
} 
\end{figure}

Last, we apply the POD-augmented WM in a channel subjected to a suddenly imposed TPG.
A schematic of the flow was shown in figure \ref{fig:sketch-chan}(b) already.
The channel is initially at a Reynolds number $Re_\tau=1000$.
The magnitude of the TPG is $10(\tau_{w,0}/\delta)$.
Table \ref{tab:SPG} shows the WMLES details.
Figure \ref{fig:SPG1}(a, b) shows the WMLES result.
We observe the following.
Firstly, the POD-augmented wall model captures the initial decrease in the streamwise wall-shear stress and gives wall-shear stress predictions that agree more closely with the DNS than the other models (comparing figure \ref{fig:SPG1}(a) and figure \ref{fig:sketch-chan}(d)).
Secondly, the streamwise wall-shear stress behaves differently from one realization to another, whereas the spanwise wall-shear stress behaves similarly between the different realizations.
Thirdly, the POD-augmented WM gives more accurate spanwise wall-shear stress predictions than the EWM.

Repeating the exercise in \S \ref{subsec:APG}, we can show that wall modeling is not critical to the prediction of the flow rate.
We can also reconstruct the velocity profiles and explain why the POD-augmented WM outperforms the EWM.
These results, however, are repetitive and therefore are not shown here for brevity.

\begin{table}%Table of simulation parameters
\begin{center}\scriptsize
\vskip -0.1in
\begin{tabular}{ccccccc}
~~~~Case~~~~    & ~~~~$Re_{\tau,0}$~~~~ & ~~~~Grid~~~~                   & ~~~~Domain~~~~                   & ~~~~TPG~~~~ & ~~~~ WM ~~~~  & ~~~~ N ~~~~\\ 
{\blue R1000T10}  & 1000          & $96\times 12\times 72$ & $8\pi\times 2\times 3\pi$ & 10 & EWM, POD-augmented WM & 10\\
\end{tabular}
\caption{\label{tab:SPG}WMLES details. The normalization is the same as in Table \ref{tab:APG}. $N$ is the number of independent realizations.
}
\end{center}
\end{table}

\section{Concluding remarks}
\label{sec:conclusions}

We augment the LoW by including an additional mode which is based on the first POD mode in a 2D channel for LES wall modeling.
The resulting wall model reconstructs the velocity in the wall layer according to
\begin{equation}
\small
    {\bf U}  ={\bf c_1} {\rm LoW}+{\bf c_2} g \, ,
    \label{eq:ensatz-PODWM}
\end{equation}
where the LoW mode includes the viscous sublayer and the buffer layer, and $g$ is the POD-based mode.
The WM and LES match at two off-wall locations instead of one so that one can solve for ${\bf c}_1$ and ${\bf c}_2$.

The results are promising.
The POD-augmented WM captures the rapid decay of the wall-shear stress when a boundary-layer flow is subjected to a large APG.
The model also captures the initial decrease in the streamwise wall-shear stress when a boundary-layer flow is subjected to a TPG.
Both phenomena had not been captured in WMLES.
{\it A-priori} analysis shows that the POD-augmented WM captures these phenomena because its ansatz is a more realistic approximation of the mean flow when there is a non-equilibrium pressure gradient.
{\blue Still, additional testing of the model in other flows will be needed to fully assess its performance.
This should include spatially developing flow, flows in complex geometries, and flows with multiple non-equilibrium effects.}

This work is the first attempt to apply results from modal analysis for predictive wall modeling in LES, and there is still a lot we can explore.
Future work will explore WMs with more modes, modes from other modal analysis techniques ({\bluenew in Appendix \ref{App:choice_of_g_mode}, we explored some other modes}), and applications of the model in flows with complex geometries.

\section*{Acknowledgment}
\noindent
C.H. and M.A. acknowledge the financial support from the Independent Research Fund Denmark (DFF) under Grant No. 1051-00015B. 
X.Y. acknowledges the Center for Turbulence Research 2022 summer program fellowship and the financial support from the Office of Naval Research under contract N000142012315. 
This work was also partially supported by the Danish e-Infrastructure Cooperation (DeiC) National HPC under grant number DeiC-AU-N2-2023005.
The authors acknowledge Michael Whitmore, Kevin Griffin, Parviz Moin, Adrian Lozano-Duran, Jane Bae, and Aurelien Vadrot for fruitful discussions.

\section*{Conflict of interest}

\noindent
The authors claim no conflict of interest.

\appendix
\section{Numerical implementation of POD}
\label{App:Num_solv_POD_eig_prob}
As we use a one-dimensional scalar variant of POD in this work the numerical implementation is surprisingly simple.
{\blue We start by collecting realizations of the fluctuating streamwise velocity component along the wall-normal direction $u'(y)$.
Here $u'(y) = u(y) - {\rm LoW}$, such that we have removed the LoW (mean).}
The realizations are collected from the DNS of a 2D turbulent channel at $Re_\tau = 5200$ \citep{lee2015direct,graham2016web}.
We sample the realizations along the wall-parallel directions and at different times as in \cite{moin1989characteristic}.
Denoting these discrete realizations as ${\bf u}_j$, we follow \cite{holmes2012turbulence} and arrange these into a data matrix

\begin{equation}
    {\bf X} = 
    \begin{bmatrix}
    {\bf u}_1 & \cdots & {\bf u}_N 
    \end{bmatrix} \, ,
\end{equation}
after which the POD analysis can be accomplished by the singular value decomposition

\begin{equation}
    {\bf X} = {\bf U} {\boldsymbol\Sigma} {\bf V}^T \, .
\end{equation}
The columns of {\bf U} are then the POD modes, while the diagonal matrix ${\boldsymbol\Sigma}$ contains the singular values (related to the POD eigenvalues), and ${\bf V}$ carries the POD mode coefficients.
For the present work, however, we are only interested in the first POD mode which is contained in the first column of ${\bf U}$.
\section{The choice of the \textbf{\textit{g}} mode}
\label{App:choice_of_g_mode}
\begin{figure}
    \centering
    \includegraphics[width=0.95\textwidth]{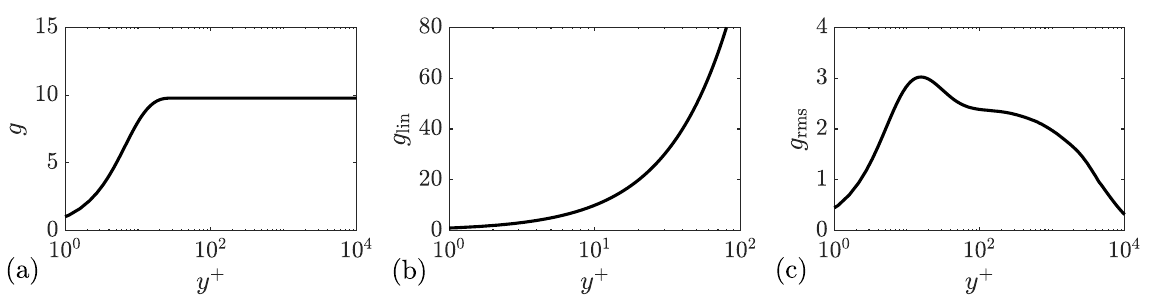}
    \caption{{\bluenew (a) The proposed POD-based $g$ mode.
    (b) The linear $g$ mode.
    (c) The rms-based $g$ mode.
    Note the different scales on both the abscissa and ordinate for better visualization of the three modes.}}
    \label{fig:different_g_modes}
\end{figure}

{\bluenew We investigate the sensitivity of the presented results to the choice of the $g$ mode.
Specifically, we consider two other choices 
\begin{equation}
    g_{\rm lin}(y^+ ) = y^+ \, , \qquad g_{\rm rms}(y^+) \propto u_{\rm rms}(y^+) \, ,
    \label{eq:new_g_modes}
\end{equation}
where $u_{\rm rms}$ is the rms of the streamwise velocity fluctuation in a $Re_\tau = 5200$ plane channel \citep{lee2015direct}.
We normalized $g_{\rm rms}$ such that $\left. dg_{\rm rms}/dy^+\right|_{y=0} = 1$. This normalization is automatically satisfied for $g_{\rm lin}$.
Additionally,  we have included an exponential term in $g_{\rm rms}$ to ensure that the mode approaches zero sufficiently far away from the wall.
This exponential term has little/no impact on the results shown below as the matching locations used here are within $y^+<1000$.
The three $g$ modes are shown in figure \ref{fig:different_g_modes}.}

\begin{figure}
\centering
\includegraphics[width=0.75\textwidth]{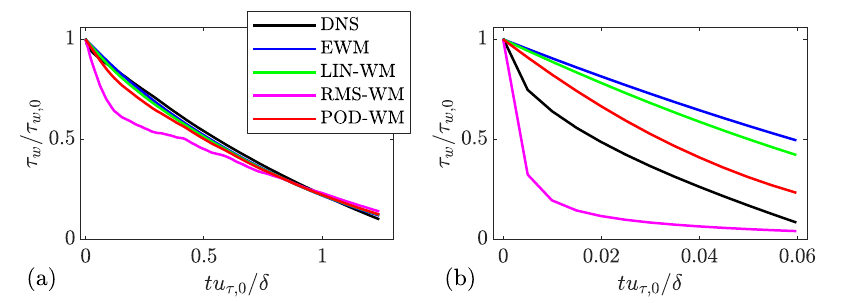}
\caption{\label{fig:WM_a_priori_APG_new} {\bluenew {\it A-priori} results for cases R1000A10 and R1000A100.
(a) Wall-shear stress from DNS, the EWM, the LIN-WM, the RMS-WM, and the POD-WM in case R1000A10.
(b) Same as (a) but for case R1000A100.}
}
\end{figure}

{\bluenew We first repeat the {\it a-priori} tests from \S \ref{subsec:non-equilib-a-priori} for cases R1000A10 and R1000A100.
The results are shown in figure \ref{fig:WM_a_priori_APG_new}.
For case R1000A10, the EWM, LIN-WM, and POD-WM yield similar predictions; 
the RMS-WM, on the other hand, predicts a low wall-shear stress within the range $0 \leq t u_{\tau,0} / \delta \leq 0.5$.
For case R1000A100, the four models all give different wall-shear stress predictions.
The LIN-WM demonstrates only a slight improvement over the EWM, with both models overestimating the wall-shear stress. In contrast, the POD-WM exhibits substantial improvement compared to the EWM and LIN-WM. Meanwhile, the RMS-WM underestimates the wall-shear stress.}

\begin{figure}
\centering
\includegraphics[width=0.75\textwidth]{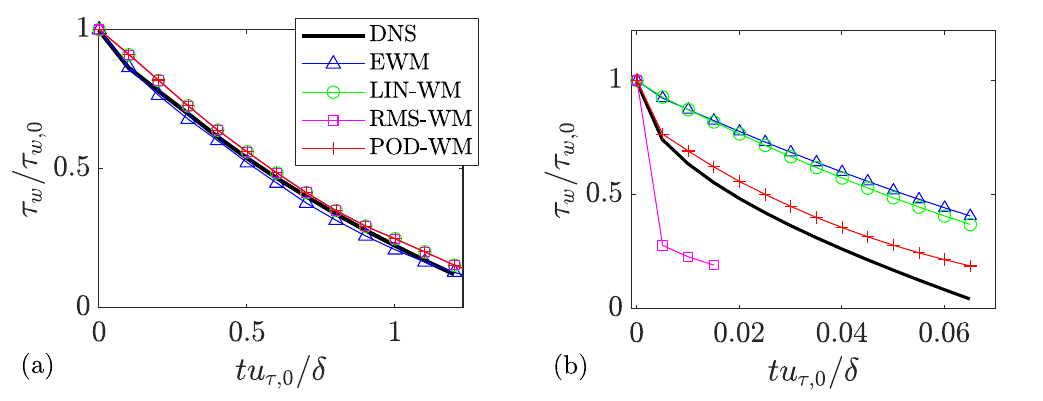}
\caption{\label{fig:WM_a_posteriori_APG_new} {\bluenew {\it A-posteriori} results for cases R1000A10 and R1000A100.
(a) Wall-shear stress from DNS, the EWM, the LIN-WM, the RMS-WM, and the POD-WM for case R1000A10.
(b) Same as (a) but for case R1000A100.
The RMS-WM result is cut off at $tu_{\tau,0}/\delta\approx 0.02$ because the WMLES simulation blows up at this point.}
}
\end{figure}

{\bluenew Next, we investigate how the {\it a-posteriori} results for the R1000A10 and R1000A100 cases, discussed in \S \ref{subsec:APG}, are affected when employing different $g$ modes specified in \eqref{eq:new_g_modes}. 
These results are presented in figure \ref{fig:WM_a_posteriori_APG_new}.
Figure \ref{fig:WM_a_posteriori_APG_new}(a) shows that the EWM, LIN-WM, RMS-WM, and POD-WM all perform well for the R1000A10 case, with their wall-shear stress predictions closely matching the DNS results.
For the R1000A100 case shown in \ref{fig:WM_a_posteriori_APG_new}(b), we observe that the EWM and LIN-WM over-predict the wall-shear stress with the LIN-WM being slightly more accurate.
The RMS-WM under-predicts the wall-shear stress by a considerable amount.
Furthermore, the simulation blows up at around $tu_{\tau,0}/\delta\approx 0.02$, when violent fluctuations in the local wall-shear stresses are found.
Among the four models, the POD-WM results are the closest to the DNS data.}

\begin{figure}
\centering
\includegraphics[width=0.75\textwidth]{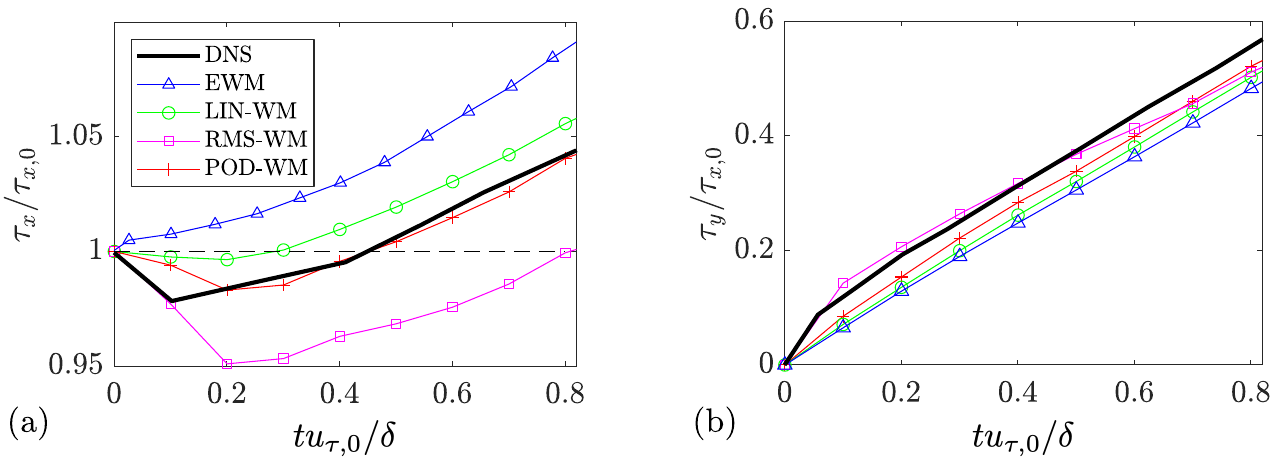}
\caption{\label{fig:WM_a_posteriori_TPG_new} {\bluenew {\it A-posteriori} results for case R1000T10.
(a) Streamwise wall-shear stress from DNS, the EWM, the LIN-WM, the RMS-WM, and the POD-WM in case R1000T10.
(b) Same as (a) but for spanwise wall-shear stress.}
}
\end{figure}

{\bluenew Finally, the TPG {\it a-posteriori} results from \S \ref{subsec:TPG}, i.e., the R1000T10 case, are revisited.
The results from the EWM, LIN-WM, RMS-WM, and POD-WM are shown in figure \ref{fig:WM_a_posteriori_TPG_new}.
For the streamwise wall-shear stress, shown in figure \ref{fig:WM_a_posteriori_TPG_new}(a), we see that the LIN-WM does not convincingly capture the initial decrease and slightly over-predicts the wall-shear stress during its increase.
The RMS-WM, on the other hand, captures the initial decrease in the streamwise wall-shear stress but under-predicts wall-shear stress for the remaining part.
Turning to the spanwise wall-shear stress in figure \ref{fig:WM_a_posteriori_TPG_new}(b), we observe that there is much less variability in the results.
The EWM, LIN-WM, and POD-WM perform similarly, with both the LIN-WM and POD-WM showing minor improvements compared to the EWM. 
The RMS-WM exhibits the closest agreement with the DNS data in the initial period.}

{\bluenew In all, we see that the choice of the $g$ mode does make a difference to the results.}

\bibliography{jfm-instructions}
\bibliographystyle{jfm.bst}
\end{document}